# Wave propagation through soils in centrifuge testing

J-F Semblat
*Laboratoire Central des Ponts et Chaussées, Paris, France*
jean-francois.semblat@lcpc.fr

M.P. Luong
*Laboratoire de Mécanique des Solides, Ecole Polytechnique, Palaiseau, France*
luong@athena.polytechnique.fr

ABSTRACT : Wave propagation phenomena in soils can be experimentally simulated using centrifuge scale models. An original excitation device (drop-ball arrangement) is proposed to generate short wave trains. Wave reflections on model boundaries are taken into account and removed by homomorphic filtering. Propagation is investigated through dispersion laws. For drop-ball experiments, spherical wave field analysis assuming linear viscoelasticity leads to a complete analytical description of wave propagation. Damping phenomena are examined and evaluated using this description.

## 1. INTRODUCTION

The dynamic behaviour of soils cannot generally be studied directly. However some special experimental methods allow direct analysis of dynamic soil behaviour [13, 15]. Nevertheless dynamic and vibratory excitations on soil often involve propagation phenomena. Such transient phenomena may prevail in the dynamic response of the medium, dealing with particular questions such as : wave dispersion, scattering and attenuation. This paper presents original wave propagation experiments in a centrifuged medium.

Wave propagation experiments in centrifuged scaled models [2, 4, 5, 10] generally concern earthquake engineering problems. Using such experiments, it is important to improve the representativeness of the source of excitation with respect to design considerations. Furthermore, the good control of wave reflection is sometimes difficult since the generated waves travel in a bounded medium of complex mechanical behaviour.

## 2. EXCITATION DEVICES

In this study, two different sources of excitation are used : an explosive earthquake simulator and a drop-ball arrangement to generate surface waves by the impact of a falling mass. The dimensions of the rigid container are the following : length L=1.30 m, width l=0.80 m and depth h=0.40 m. It is filled with dry Fontainebleau sand of unit weight $\gamma$=16.5 kN.m$^{-3}$ and mean grain diameter 196 µm.

### 2.1 Earthquake simulations

Earthquake simulation devices are often used to study soil-structure interaction problems. These generally consist of an explosion room coupled with a mechanical filter. The pressure wave created is transmitted to the soil mass by a flexible membrane. This kind of experimental device presents two important drawbacks :
- the long duration of the generated wave train makes wave reflection phenomena difficult to detect and to circumvent ;
- the excitation is not directly transmitted to the soil mass : there is a container-excitation interaction (part of the excitation travels in the container itself).

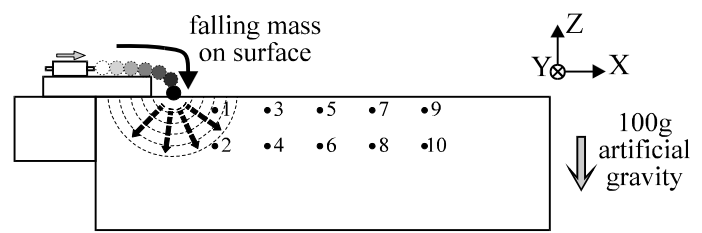

Figure 1 : *Drop-ball arrangement for dynamic excitation on soil.*

### 2.2 Drop-ball arrangement

Classical devices for wave propagation analysis in a centrifuged medium (explosive sources, vibratory loadings...) do not generally allow an easy control of the generated wave train (duration, transmission...). Our purpose is to generate a wave field of short duration using a simple experimental system. As it is shown in Figure 1, a pneumatic actuator pushes a spherical ball which then falls onto the medium surface. All sensors have the same coordinate along the Y-axis. The whole experimental device is called « drop-ball arrangement ». It gives both short excitation duration and direct loading transmission [13,14].

### 2.3 Acceleration measurements

Three-dimensional acceleration measurements are performed at ten different locations positionned at two different depths (see Figure 1). Examples of horizontal acceleration signals (X-axis) are given in Figures 2 and 3. For earthquake simulations, the spectral content is rich, but the signal duration is long compared with the propagation delays through the container (see Figure 2 and [13]). For the drop-ball arrangement (Figure 3), acceleration signal duration is much shorter : it allows a good control of propagation phenomena and a precise analysis of dispersion and the attenuation process.

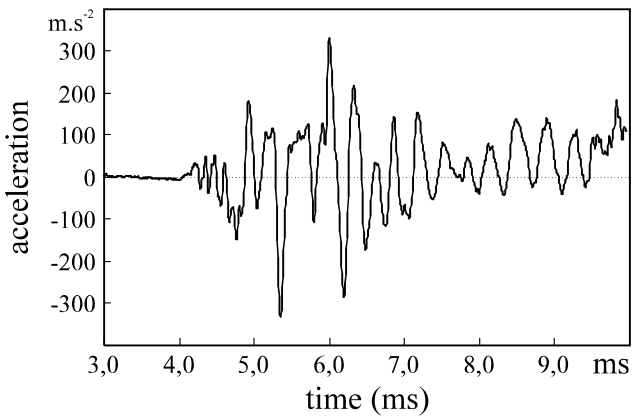

Figure 2 : *Acceleration signal for explosive earthquake simulations.*

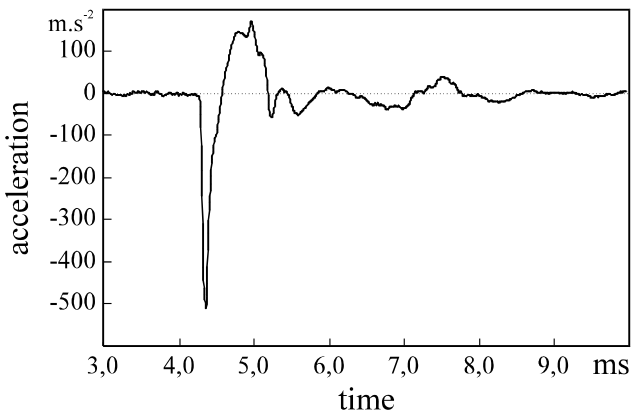

Fig.3: *Acceleration signal for drop-ball experiments.*

For drop-ball experiments, as the generated wave field is spherical, distances from the source to the sensors have to be calculated. Table I gives values of all the source-sensor distances.

| surface sensors | | | | | |
|---|---|---|---|---|---|
| number | 1 | 3 | 5 | 7 | 9 |
| distance | 0.11 m | 0.24 m | 0.37 m | 0.50 m | 0.63 m |
| mid-depth sensors | | | | | |
| number | 2 | 4 | 6 | 8 | 10 |
| distance | 0.19 m | 0.28 m | 0.40 m | 0.52 m | 0.65 m |

Table I : *Distance from the spherical wave source to the sensors.*

### 2.4 Waves characterization

From the 3D measurements, it is possible to determine the acceleration vectors. For earthquake simulations, they are difficult to build because of the fast and irregular changes in their directions and magnitudes. For drop-ball experiments, acceleration vectors are quite easy to determine and to analyse because there is no interaction between the direct and reflected waves.

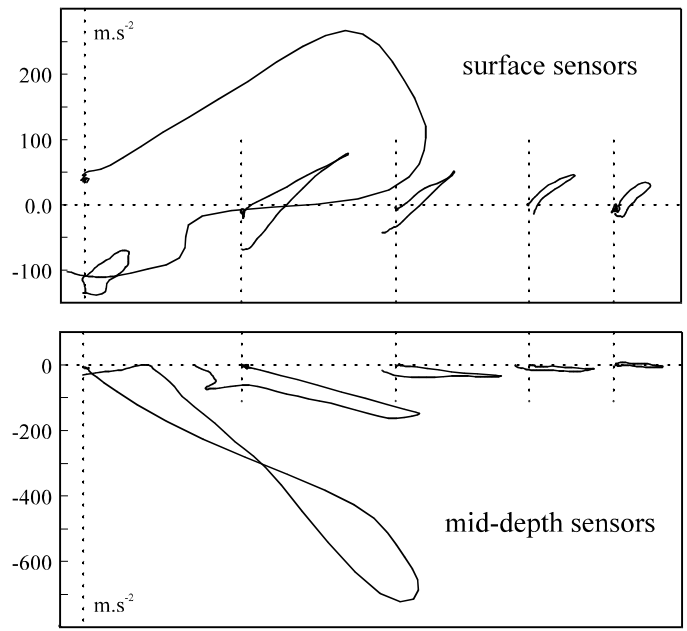

Figure 4 : *Accelerations in XZ plane from experimental results.*

Figure 4 gives acceleration values in the XZ plane for a drop-ball experiment : surface sensors at top (odd numbers) and mid-depth sensors at bottom (even numbers). It is then possible to determine maximum accelerations and to build the corresponding acceleration vectors in XZ plane (acceleration along the Y axis is negligible).

For each sensor, the maximum acceleration magnitude and corresponding vector angle are estimated from the results of Figure 4. The angle between the vector direction and the horizontal axis X is denoted $\alpha_i$ for *sensor « i »*. Acceleration vectors are given in Figure 5 for all sensors.

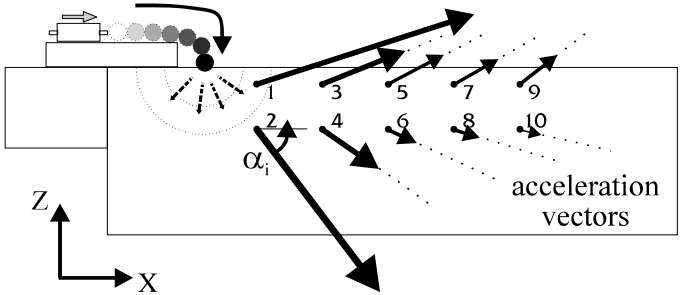

Figure 5 : *Acceleration vectors in the XZ plane : pressure waves at mid-depth and different wave types for near-surface sensors.*

Incident angles $\beta_i$ give the direction of the line between the point of impact and the location of sensor *i*. The comparison between angles $\alpha_i$ and incident angles $\beta_i$ shows an obvious coincidence between the incidence directions and acceleration vectors directions for mid-depth sensors ($\alpha_i=\beta_i$ for i=2, 4, 6, 8 and 10). For drop-ball experiments, the acceleration waves generated at mid-depth are clearly identified as pressure waves. This is a very important conclusion for the study of attenuation phenomena.

## 2.5 Some preliminary results
*Wave velocity*
Estimation of wave velocity is firstly made directly from peak delay measurements on time signals or using different signal processing methods [13] (crosscorrelation functions, time phase). A more precise evaluation of wave velocities (phase and group velocities) is subsequently proposed using time-frequency analysis of acceleration signals.

Values reached for explosive earthquake simulations lead to a mean wave velocity : $V_{mean}$=500 m.s$^{-1}$ [13]. For drop-ball experiments, values of wave velocities corresponding to different pairs of sensors are given in Table II (test n°13). The mean velocity is estimated to be : $V_{mean}$=423 m.s$^{-1}$. Note that there is no significant difference between surface and mid-depth velocities (due to initial homogeneity of the medium). Wave velocities are thus nearly constant in the whole centrifuged model. For drop-ball experiments, they are slightly lower than those measured during explosive earthquake simulations.

|  | wave velocities | | | |
|---|---|---|---|---|
| surface | 1/3 | 3/5 | 5/7 | 7/9 |
|  | 433 m/s | 400 m/s | 426 m/s | 448 m/s |
| mid-depth | 2/4 | 4/6 | 6/8 | 8/10 |
|  | 409 m/s | 421 m/s | 429 m/s | 419 m/s |

Table II : *Wave velocities for drop-ball test n°13*

*Simple numerical results*
For drop-ball experiments, the generated wave field is assumed to be spherical. The numerical model is then chosen axisymetrical. Several numerical results (linear elasticity) are presented in Figure 6. They give interesting qualitative features of wave propagation phenomena in the centrifuged model : wave type, wave reflection on medium boundaries... These numerical computations are performed using civil engineering dedicated finite element system CESAR-LCPC. Numerical results given in Figure 6 correspond to four different time values :

- *time $T_1$* : wave propagation begins, the different wave types cannot be distinguished ;
- *time $T_2$* : pressure waves (the fastest ones) are well distinguished from shear waves and surface waves ;
- *time $T_3$* : reflection of pressure waves at the bottom of the container ;
- *time $T_4$* : direct pressure and shear waves keep travelling whereas the reflected P wave is propagated to the free surface.

From experimental results given in Figure 4 and 5, mid-depth sensors detect only pure pressure waves. As shown by experimental and numerical results (Figures 4, 5 and 6), different wave types appear near the free surface : shear waves, surface waves. In the following sections (§5 et 6), analytical models are proposed to study and precisely quantify attenuation and dispersive phenomena.

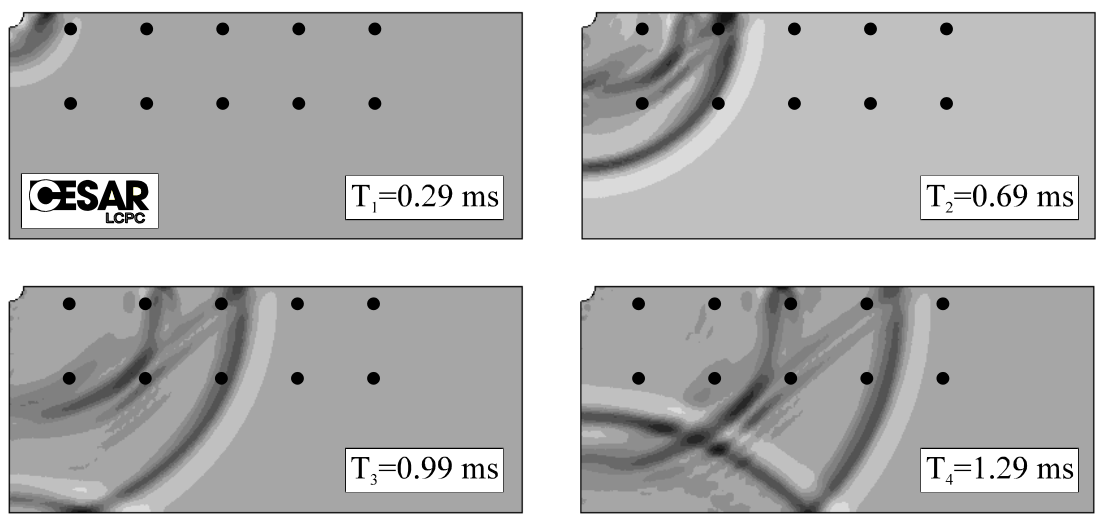

Figure 6 : *Numerical modelling of spherical wave propagation (CESAR-LCPC).*

## 2.6 Scale factors for dynamic problems

Centrifuge experiments are a realistic way of testing materials and structures, as stresses are identical in the centrifuge scaled model despite its dimensions. This is made possible thanks to artificial gravity forces allowing the scaling relations to be fulfilled. These scaling relations between full-scale structure and scaled model are determined considering physical relations characterising the problem (balance equations, constitutive relations...).

| mechanical parameter | scale factor | numerical value |
|---|---|---|
| length | $l^*$ | 1/100 |
| time | $t^*$ | 1/100 |
| velocity | $v^*$ | 1 |
| frequency | $f^*$ | 100 |
| acceleration | $a^*$ | 100 |
| viscosity | $h^*$ | 1/100 |
| damping | $a^*$ | 100 |

Table III : *Scaling relations for dynamic problems*

In the experiments performed, sand is centrifuged under a 100$g$ gravity (where $g$ is 9.81 m.s$^{-2}$) and scaling relations for length, time, velocity are given in Table III. For dynamic testing, it is necessary to meet scaling relations considering transient excitation and propagation phenomena. Scaling relations have to involve acceleration magnitude, frequency and also viscosity and damping for attenuation phenomena. These quantities could be usually evaluated by conventional resonnant tests.

## 3. REFLECTED WAVES REMOVAL BY HOMOMORPHIC FILTERING

### 3.1 Wave reflections in drop-ball experiments

One of the main advantages of drop-ball experiments is the generation of short duration excitation. The detection and control of wave reflections are thereby made easier. As can be seen from Figure 7, for drop-ball experiments, direct and reflected waves appear in the acceleration signals.

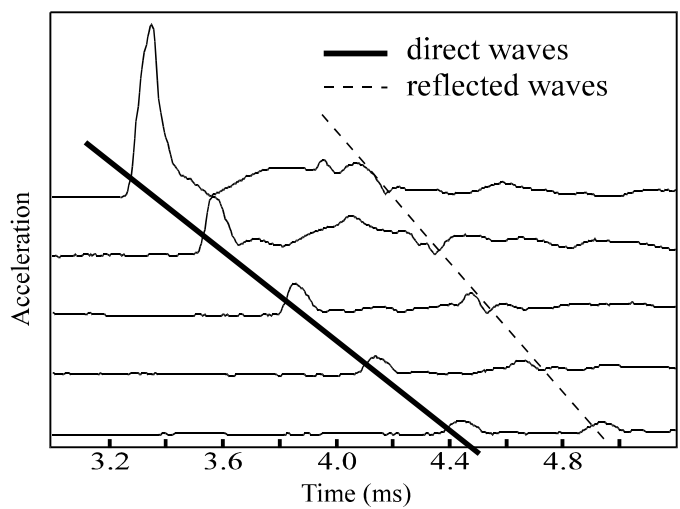

Figure 7 : *Acceleration measurements (magnitude) for mid-depth sensors showing direct and reflected waves (drop-ball test).*

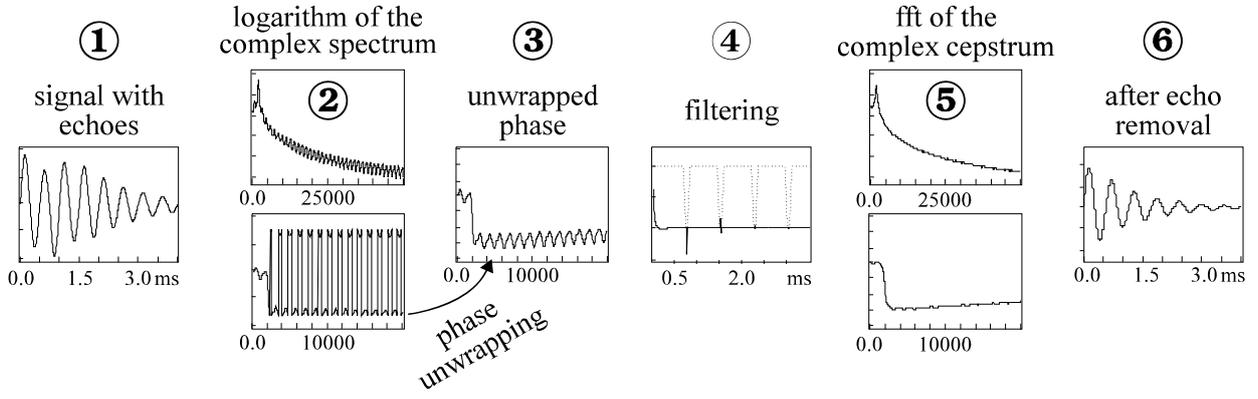

Figure 8 : *Complex cepstrum calculation procedure on a synthetic signal*

The detection of reflected waves is thus quite easy. It is then possible to estimate the delays beween the direct and reflected waves (at the bottom of the container). Estimated experimental delays and calculated theoretical delays (assuming straight rays) are in agreement for all sensors [13]. The theoretical assumption considering straight rays is consequently acceptable. Furthermore, it would be interesting to remove reflected waves from the measured acceleration signal. For drop-ball experiments, it is not very difficult thanks to the short duration of the excitation. In a more general case, the method proposed in the following section (homomorphic filtering) works for every kind of excitation, even if it is of long duration.

### 3.2 Homomorphic filtering
*3.2.1 Main principles*
Homomorphic signal processing is based on a generalised linear systems theory [18, 19]. Homomorphic processing allows source excitation determination avoiding problems of echoes or wave reflection. This technique was used by Ulrych [19] to study attenuation and elastic wave dispersion by separating source terms from propagation terms.

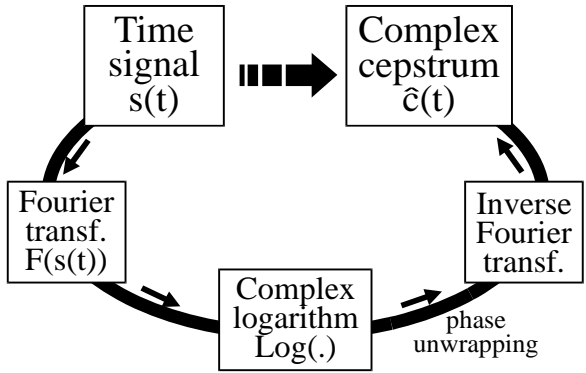

Figure 9 : *Main stages of homomorphic filtering.*

*3.2.2 Complex cepstrum calculation*
Complex cepstrum $\hat{c}(\tau)$ is calculated from magnitude and phase spectra. Figure 9 shows the main calculation stages. If $F$ denotes the Fourier transform and $Log$ the complex logarithm, the complex cesptrum $\hat{c}(\tau)$ of the signal $s(t)$ can be written :

$$\hat{c}(\tau) = F^{-1}\left[Log(F(s(t)))\right]$$

Hopelessly the phase spectrum $Arg[S(f)]$ is generally wrapped and must be converted to an unwrapped form. J.Tribolet [17] proposed a special unwrapping algorithm based on the phase derivative calculation. Its theoretical expression is the following :

$$\frac{d(Arg[S(f)])}{df} = \frac{S_R(f).S_I'(f) - S_I(f).S_R'(f)}{|S(f)|^2}$$

where $S(f)$ is the complex spectrum of $s(t)$. From this expression, it is possible to determine the complex cepstrum $\hat{c}(\tau)$. The different stages of the procedure are presented in Figure 8 for a synthetic signal. Phase unwrapping (stage 3) precedes the real filtering stage.

The complex cepstrum shows peaks characterizing echoes or reflected wave delays. To remove reflected waves, peaks in the complex cepstrum are suppressed by a weighting window (stage 4 of Figure 8). The « inverse cepstral transform » will then yield the original signal without any reflected waves [13, 18, 19].

*3.2.3 Application to drop-ball experiments*
Figure 10 gives an acceleration signal on which waves reflected at the bottom of the container are easy to detect (thin line). The same signal after homomorphic filtering is entirely free of any reflection (thick line).

This deconvolution technique is very useful for acceleration signals in a centrifuge. This method is also efficient for long-duration signals (see Figure 8) : even if direct and reflected waves are thoroughly mixed, homomorphic filtering can separate both (or more) wave trains. It is a good alternative to isolation techniques (see J.A.Cheney et al. [4]).

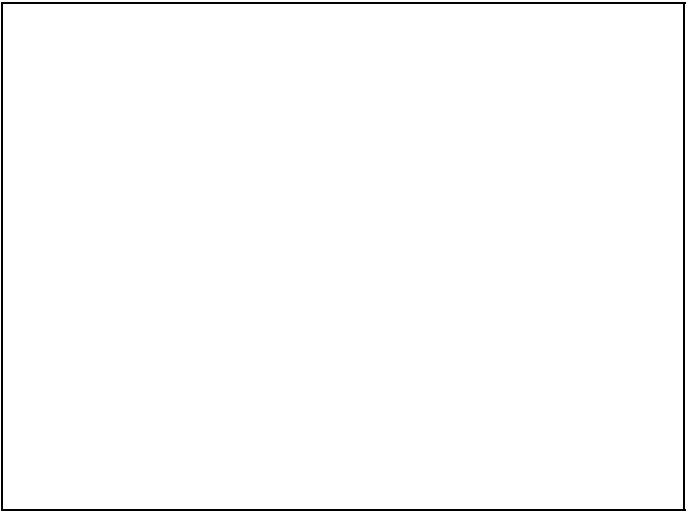

Figure 10 : *Reflected wave removal by homomorphic filtering (drop-ball arrangement).*

## 4. ANALYSIS OF DISPERSION

Once reflection phenomena are detected and well controlled (for drop-ball experiments), a complete analysis of propagation calls for the study of wave dispersion in the medium. Signal processing by time-frequency analysis is proposed for that purpose.

### 4.1 Time-frequency analysis

*4.1.1 Purpose*

The decomposition of the acceleration signals in the time-frequency domain gives dispersive characteristics of propagation [6, 7, 8, 16]. This analysis can be performed using different methods : filtering, wavelet transform... In the different frequency bands, dispersion laws (phase and group velocities) of the centrifuged medium are then determined.

*4.1.2 Wigner-Ville distribution*

The Wigner-Ville distribution gives an overall theoretical approach to the time-frequency analysis of signal [7]. The main advantage of this method is that the choice of the weighting functions in the time and frequency domains is free. Even for strongly dispersive media, the Wigner-Ville distribution allows the determination of group velocity. Sessarego et al. used this method to study the dispersion of Lamb waves in a duraluminum plate [16].

*4.1.3 Wavelet transform*

The Wigner-Ville distribution includes many other types of transforms : shifted Fourier transform, wavelet transform, etc... The wavelet transform is built using a family of elementary functions $\psi_{ab}(t)$ defined from an analysing wavelet $\psi(t)$ [7, 16]. This family is expressed under the following form :

$$\psi_{ab}(t) = \frac{1}{\sqrt{a}} \cdot \psi\left(\frac{t-b}{a}\right), \qquad b \in \Re, a > 0$$

Wavelet coefficients of the signal *s* allow the characterization of *s* in both time and frequency domains. They are written in the following form :

$$C_s(a,b) = \langle s, \psi_{ab} \rangle = \int_{-\infty}^{+\infty} s(t) \cdot \overline{\psi}_{ab}(t) \cdot dt$$

Changing parameters a and b, the analysing wavelet $\psi(t)$ is simultaneously translated (parameter b) and expanded (or contracted, parameter a). The signal may then be analysed in the time-frequency domain after decomposition on the generated wavelet family.

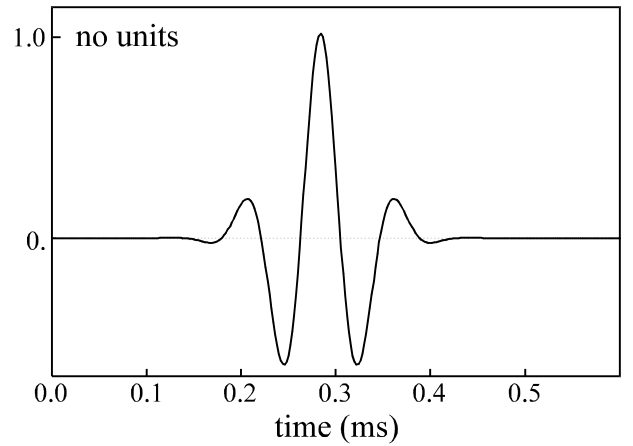

Figure 11 : *Analysing wavelet used for the time-frequency analysis of acceleration signals.*

An example of the analysing wavelet used in this study is given in Figure 11. One hundred wavelets are built from this analysing wavelet to determine $C_S(a,b)$ coefficients. Dispersive phenomena are then studied between 500 to 9500 Hz (see also [13]).

### 4.2 Analysis of dispersion

Acceleration signals are analyzed in different frequency bands (see some examples in Figure 12), and phase delays are estimated from these analyzed signals. The determination of group delays is made from the envelope curves of the signals [6, 8]. From phase and group delays, phase and group velocities are easily obtained [6, 8, 13].

The 3D-diagram presented in Figure 13 gives envelope curves of acceleration signals filtered in different frequency bands. From this diagram, it is possible to determine group delays as a function of frequency. The thick line gives group delays for different frequencies.

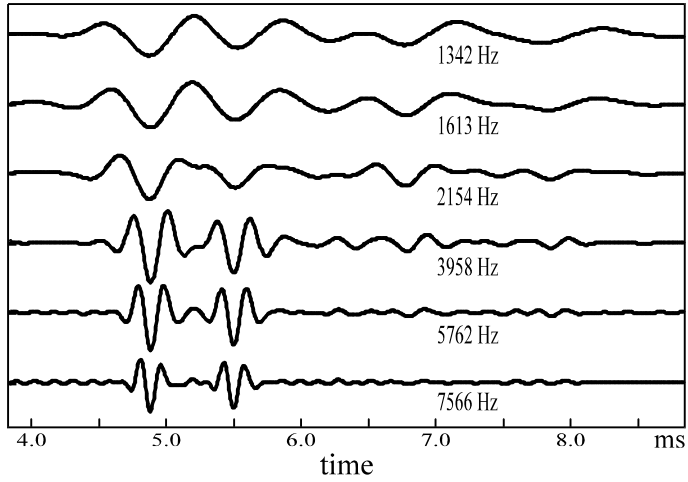

Figure 12 : *Filtered signals for different frequency bands (drop-ball).*

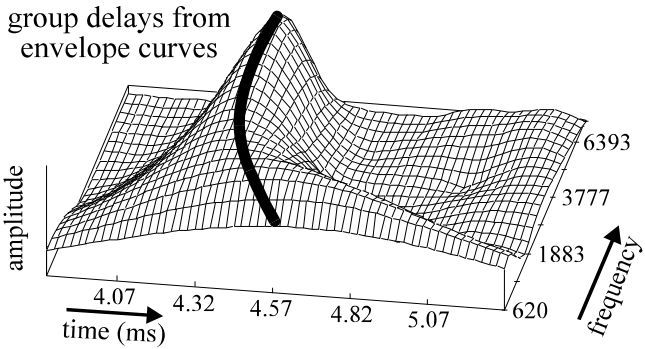

Figure 13 : *3D-diagram giving envelope curves for different frequency bands (drop-ball experiments).*

For drop-ball experiments, wave reflection phenomena are easy to detect. As can be seen in Figure 12, direct and reflected waves are both present in filtered signals for medium and high frequencies, but for frequencies under 1500Hz, there is just one wave train. Actually, for low frequencies, the wavelengths are higher than the container dimensions and there is a strong geometrical dispersive effect. This can be seen in Figure 13 : group delays increase sharply below 1500 Hz :
- for low frequencies, there is a strong geometrical dispersive effect (waveguide effect). Group and phase velocities are largely underestimated (group and phase delays being overestimated) ;
- for medium and high frequencies, there is a material dispersive effect (weaker than the previous geometrical one). Reflected waves are not mixed with direct waves in the filtered signals (see Figure 12) : phase and group velocities are correctly estimated.

| frequency (in Hz) | drop-ball test n° 12 (velocities in m.s$^{-1}$) | | drop-ball test n° 13 (velocities in m.s$^{-1}$) | |
|---|---|---|---|---|
| | $V_\Phi$ | $V_g$ | $V_\Phi$ | $V_g$ |
| 440 | 350.8 | 266.5 | 353.2 | 267.4 |
| 975 | 389.7 | 433.4 | 368.3 | 453.8 |
| 1710 | 429.4 | 413.2 | 431.0 | 396.8 |
| 3175 | 427.3 | 424.3 | 422.9 | 420.1 |
| 5320 | 416.9 | 405.1 | 418.4 | 411.7 |
| 9280 | 415.4 | 416.5 | 418.4 | 426.9 |

Table IV : *Experimental values of phase and group velocities.*

Table IV gives experimental values of phase and group velocities for two drop-ball experiments and in six different frequency bands. These experimental values are compared further with theoretical phase and group velocities.

## 5. ATTENUATION

In this section, attenuation phenomena are investigated and related to the analysis of dispersion presented above with a viscoelastic formulation.

### 5.1 Geometrical attenuation

For drop-ball experiments, the generated wave field is spherical. There must be a strong geometrical contribution to the attenuation process which is generally considered (for an infinite medium) to be inversely proportional to the distance from the source of excitation. The acceleration wave magnitude at point $M_i$ can then be written as : $A_i = A_0/r_i$. The comparison of acceleration magnitudes at two points $M_i$ and $M_j$ leads to the following expression :

$$A_j = A_i \cdot \left( \frac{r_i}{r_j} \right)$$

### 5.2 Material attenuation

Centrifuge experiments performed in the 80's by Luong suggested a physical (or material) attenuation of the form $a(x) = a_0 \cdot exp(-\alpha \cdot x)$, where the damping coefficient $\alpha$ is assumed to be independent of frequency and equal to 1,1 m$^{-1}$. For drop-ball experiments, combining the geometrical and material attenuations, the comparison of acceleration magnitudes for two different points is of the following form :

$$A_j = A_i . \exp\left[-\alpha.(r_j - r_i)\right].\frac{r_i}{r_j} \quad (1)$$

This expression allows the estimation of the acceleration signal at a point $M_j$ from a measured signal at point $M_i$. The ratio $r_i/r_j$ takes into account the geometrical attenuation (wave front expansion), whereas *exp[-α.(r$_j$-r$_i$)]* gives the material attenuation between $M_i$ and $M_j$. For drop-ball experiments, this expression does not correctly describe the attenuation phenomena [13].

**5.3 Estimation from experimental results**

From the previous description of attenuation, the amplitude reduction factor, estimated from experimental results, can be compared with the geometrical attenuation. The amplitude reduction factor is drawn from acceleration magnitudes in the XZ plane, and geometrical attenuation (for infinite medium and spherical wave field) is determined by the distance ratio $r_i/r_j$ (where i and j are the sensors numbers). From the amplitude reduction factor (related to total attenuation), it is then possible to compare geometrical and material attenuation. Table V gives values for the three amplitude reduction factors : the experimental (total), the geometrical and the material one. Material attenuation is the ratio of total and geometrical attenuations (corresponding to the exponential term in expression (1)). Near the source, acceleration amplitude is the highest for mid-depth sensors whereas, further away from the source, surface amplitude is slightly higher.

Experimental and geometrical amplitude reduction factors are estimated from maximum acceleration values and sensors distances (see Table V). Experimental amplitude reduction factors values are fairly similar for surface and mid-depth senors. However, geometrical attenuation is higher for surface sensors near the source (low reduction factor) and the related material attenuation is thus lower for surface sensors. Nevertheless, it should be noted that expression (1) is best adapted to mid-depth accelerations (and not really to surface acceleration waves) and non linear effects are possibly involved near the source.

For mid-depth sensors, the geometrical amplitude reduction factor increases with distance (geometrical attenuation accordingly decreases). From Table V, it appears that there is no material attenuation at all for the last two mid-depth sensors.

| sensor number | sensor distance | acceleration magnitude | total amplitude reduction | geometrical attenuation | material attenuation |
|---|---|---|---|---|---|
| 1 | $r_1$=0.11 m | $a_1$=757.11 m.s$^{-2}$ | $a_3/a_1$=0.326 | $r_1/r_3$=0.458 | 0.712 |
| 3 | $r_3$=0.24 m | $a_3$=246.76 m.s$^{-2}$ | $a_5/a_3$=0.556 | $r_3/r_5$=0.649 | 0.857 |
| 5 | $r_5$=0.37 m | $a_5$=137.15 m.s$^{-2}$ | $a_7/a_5$=0.810 | $r_5/r_7$=0.740 | 1.09 |
| 7 | $r_7$=0.50 m | $a_7$=111.13 m.s$^{-2}$ | $a_9/a_7$=0.767 | $r_7/r_9$=0.794 | 0.966 |
| 9 | $r_9$=0.63 m | $a_9$=85.24 m.s$^{-2}$ | | | |
| 2 | $r_2$=0.19 m | $a_2$=877.96 m.s$^{-2}$ | $a_4/a_2$=0.349 | $r_2/r_4$=0.679 | 0.514 |
| 4 | $r_4$=0.28 m | $a_4$=306.45 m.s$^{-2}$ | $a_6/a_4$=0.544 | $r_4/r_6$=0.700 | 0.777 |
| 6 | $r_6$=0.40 m | $a_6$=166.62 m.s$^{-2}$ | $a_8/a_6$=0.601 | $r_6/r_8$=0.769 | 0.782 |
| 8 | $r_8$=0.52 m | $a_8$=100.08 m.s$^{-2}$ | $a_{10}/a_8$=0.744 | $r_8/r_{10}$=0.800 | 0.930 |
| 10 | $r_{10}$=0.65 m | $a_{10}$=74.50 m.s$^{-2}$ | | | |

Table V : *Comparison of amplitude reduction factors (total, geometrical, material).*

## 6. SIMULATIONS WITH RHEOLOGICAL MODELS

### 6.1 Simplified solution for drop-ball tests

*6.1.1 Equation of motion*

The analysis of three-dimensional acceleration measurements shows that, for mid-depth sensors, acceleration waves are pressure waves (see Figure 5). We assume that the wave field is spherical and keep only the radial component of the field variables. The equation of motion, consequently, takes the following form :

$$div(\sigma(r,t)) = \frac{\partial \sigma_{rr}(r,t)}{\partial r} + \frac{2}{r}.\sigma_{rr}(r,t) = \rho.\frac{\partial^2 u_r(r,t)}{\partial t^2} \quad (2)$$

This differential equation can be rewritten in a simpler form using Fourier tranforms of both radial stress and displacement ($\sigma_{rr}$ and $u_r$). The Fourier transforms (or complex spectra) of $\sigma_{rr}(r,t)$ and $u_r(r,t)$ are denoted $\sigma_{rr}^*(r,\omega)$ and $u_r^*(r,\omega)$.

*6.1.2 Viscoelastic models*

To solve equation (2), a constitutive model for the medium is needed. Linear viscoelasticity gives a simple relation beween stress and strain in the frequency domain. It is not realistic near the source but further away from the source it could be a good approximation of damping phenomena. The three viscoelastic models presented in Figure 14 lead to the following relation :

$$\sigma_{rr}^*(r,\omega) = E^*(\omega).\varepsilon_{rr}^*(r,\omega) \quad (3)$$

where $E^*(\omega)$ is the complex modulus. The different forms taken by this complex modulus are given in Figure (14). The corresponding variations of attenuation versus frequency are shown in Figure (15).

The equation of motion (2) combined with the constitutive equation (3) admits the following solution, in terms of radial displacement :

$$u^*(r,\omega) = \frac{U_0(\omega)}{r}.\exp(i.\xi(\omega).r) \quad (4)$$

where $\xi$ is the complex function of $\omega$ defined by :

$$\xi^2(\omega) = \frac{\rho.\omega^2}{E^*(\omega)} \quad (5)$$

in which $\xi(\omega)$ is the complex wave number. Its real part corresponds to the phase shift and its imaginary part to the damping coefficient [13]. In this expression, $U_0(\omega)$ is the source term independent of $r$.

Attenuation is expressed through two different parameters :
- geometrical attenuation due to spherical wave field propagation and inversely proportional to $r$ (distance from the source) but independent of frequency ;
- viscous damping given by the imaginary part of $\xi(\omega)$ and frequency-dependent.

Expression (4) completely characterises the spherical (pressure) wave field in linear viscoelastic medium. It is then possible to simulate the propagation of these pressure waves for mid-depth sensors.

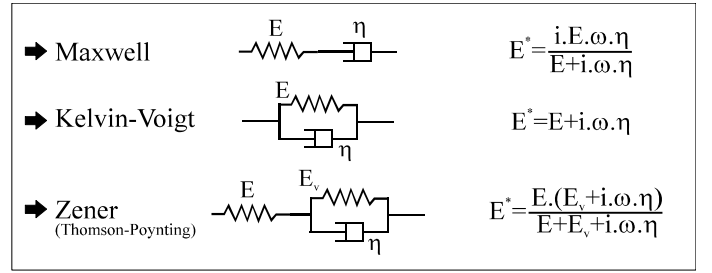

Figure 14 : *Viscoelastic models used for simulations and the corresponding complex modulus.*

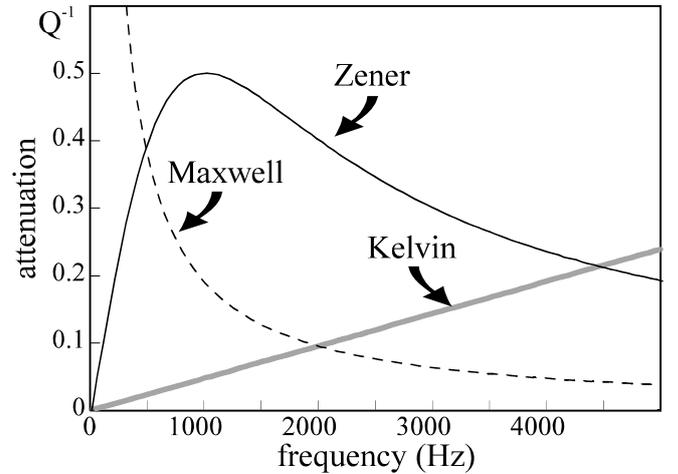

Figure 15 : *Attenuation versus frequency for different viscoelastic models.*

### 6.2 Complete solution (P-wave)

*6.2.1 Stress and strain tensors*

Previous assumption reduces strain tensor to only the first diagonal term $\varepsilon_{rr}$. Considering a purely radial displacement (as is verified experimentally), the complete strain tensor can be written :

$$\underline{\underline{\varepsilon}} = \begin{bmatrix} \frac{\partial u_r}{\partial r} & \frac{1}{2r}\frac{\partial u_r}{\partial \theta} & \frac{1}{2r.\sin\theta}\frac{\partial u_r}{\partial \theta} \\ \frac{1}{2r}\frac{\partial u_r}{\partial \theta} & \frac{u_r}{r} & 0 \\ \frac{1}{2r.\sin\theta}\frac{\partial u_r}{\partial \theta} & 0 & \frac{u_r}{r} \end{bmatrix}$$

Stress tensor is then determined assuming linear viscoelasticity :

$$\sigma_{rr} = \left(\lambda^* + 2\mu^*\right)\frac{\partial u_r}{\partial r} + \lambda^* \frac{2u_r}{r}$$

$$\sigma_{\theta\theta} = \lambda^*\left(\frac{\partial u_r}{\partial r} + \frac{2u_r}{r}\right) + 2\mu^* \frac{u_r}{r}$$

$$\sigma_{\varphi\varphi} = \lambda^*\left(\frac{\partial u_r}{\partial r} + \frac{2u_r}{r}\right) + 2\mu^* \frac{u_r}{r}$$

$$\sigma_{r\theta} = \frac{\mu^*}{r}\frac{\partial u_r}{\partial \theta} \ ; \ \sigma_{r\varphi} = \frac{\mu^*}{r.\sin\theta}\frac{\partial u_r}{\partial \varphi} \text{ and } \sigma_{\theta\varphi} = 0$$

where $\lambda^*$ and $\mu^*$ are complex Lamé constants.

### 6.2.2 Solution of the complete problem

No dependence in $\theta$ and $\varphi$ (second and third spherical coordinates) is taken into account. Equation of motion leads to the following differential equation :

$$\frac{\partial^2 u_r}{\partial r^2} + \frac{2}{r}\frac{\partial u_r}{\partial r} - \frac{2}{r^2}u_r = \frac{\rho}{\lambda^* + 2\mu^*}\frac{\partial^2 u_r}{\partial t^2} \quad (6)$$

The solution of the differential equation (6) is then :

$$U^*(r,\omega) = \frac{U_0^*(\omega)}{r}.\left(\xi(\omega) + \frac{i}{r}\right).exp(i.\xi(\omega).r) \quad (7)$$

It is slightly different from the simplified solution (4). It involves a term proportional to $1/r^2$ decreasing rapidly with increasing distance r. Differences between these two solutions are discussed in the next paragraph.

### 6.3 Simulations for drop-ball experiments

### 6.3.1 Other assumptions

*Transient wave calculation*

The measurements performed in centrifuge experiments concern acceleration parameters. Expression (4) is also verified by the Fourier transform of acceleration signals $a^*(r,\omega)$. This expression is verified for each frequency component. Using simplified solution (4)), acceleration $a(r,t)$ in the time domain is obtained by inverse Fourier transformation under the following form :

$$a(r,t) = \frac{1}{r}.\int_{-\infty}^{+\infty} A_0(\omega).exp(i.\xi(\omega).r).exp(i.\omega.t).d\omega$$

*Behaviour parameters*

For Maxwell and Kelvin-Voigt models, the behaviour parameters are Lamé constants $\lambda$ and $\mu$ and viscosity coefficient $\eta$. $\lambda$ and $\mu$ can be easily calculated from pressure waves velocity $C$ using the relation :

$$C = \sqrt{\frac{\lambda + 2\mu}{\rho}} \quad \text{thus} \quad \lambda + 2\mu = \rho.C^2$$

where $\rho$ is the density of the sand ($\rho$=1650 kg/m$^3$) and C=423 m.s$^{-1}$ for drop-ball experiments (see §2.5).

### 6.3.2 Simulated acceleration

*Simplified solution*

Simulations were made for mid-depth sensors to take into account only the pressure waves. Considering the acceleration spectra for two sensors, $i$ and $j$, the ratio of these two spectra can easily be written in frequency domain from expression (4). Denoting $a_m^*(r,\omega)$ as the Fourier transform of acceleration, this ratio takes the following form :

$$\frac{a_m^*(r_j,\omega)}{a_m^*(r_i,\omega)} = \frac{r_i}{r_j}.exp\left[i.\xi(\omega).(r_j - r_i)\right] \quad (8)$$

If acceleration is known at distance $r_i$ from the source, acceleration at distance $r_j$ can be determined from viscoelastic simulations. From expression (8), the simulated acceleration $a_m^{sim}(r_j,t)$ can be expressed, after inverse Fourier transformation, as :

$$a_m^{sim}(r_j,t) = \int_{-\infty}^{+\infty} a_m^*(r_i,\omega).\frac{r_i}{r_j}.exp\left[i.\xi(\omega).(r_j - r_i)\right]$$
$$.exp(i.\omega.t).d\omega \quad (9)$$

*Simulations for the complete solution*

The complete solution defined by equation (7) gives the relation between the measured acceleration at distance r$_i$ and the simulated acceleration at distance r$_j$. This solution can be written as follows :

$$\frac{a_m^*(r_j,\omega)}{a_m^*(r_i,\omega)} = \frac{r_i^2}{r_j^2}.\frac{r_j.\xi(\omega) + i}{r_i.\xi(\omega) + i}.exp\left[i.\xi(\omega).(r_j - r_i)\right]$$

This expression involves a term proportional to the ratio between squared distance. This term must decrease very quickly with increasing distance. It appears clearly in Figure 16 : near the source, the simplified solution and the complete solution lead to very different simulations. For an increasing source distance, the difference between both solutions becomes small (see Figure 16).

From the mechanical parameters determined, the difference between simplified and complete solutions (equations (4) and (7)) becomes small when the distance to the source is greater than 0.05m. As is the case for all accelerometers, the simplified solution is chosen.

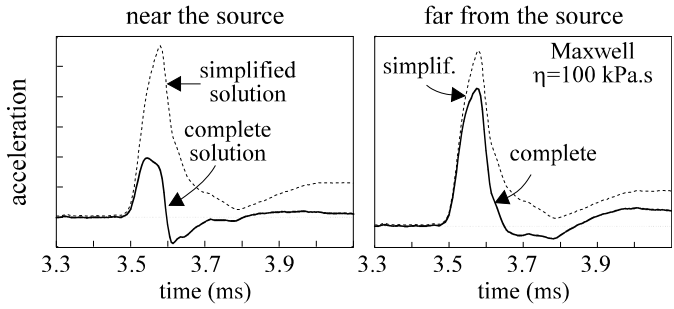

Figure 16 : *viscoelastic modelling : comparison between simplified and complete solutions.*

### 6.3.3 Simulations from experimental results

For the Maxwell and Kelvin-Voigt models, values of $\eta$ have to be determined to agree with experimental results. This agreement is evaluated from the first acceleration peak. The best $\eta$ values empirically determined by this procedure are the following :

- $\eta$=150,000 Pa.s for Maxwell's model
- $\eta$=1000 Pa.s for Kelvin's model

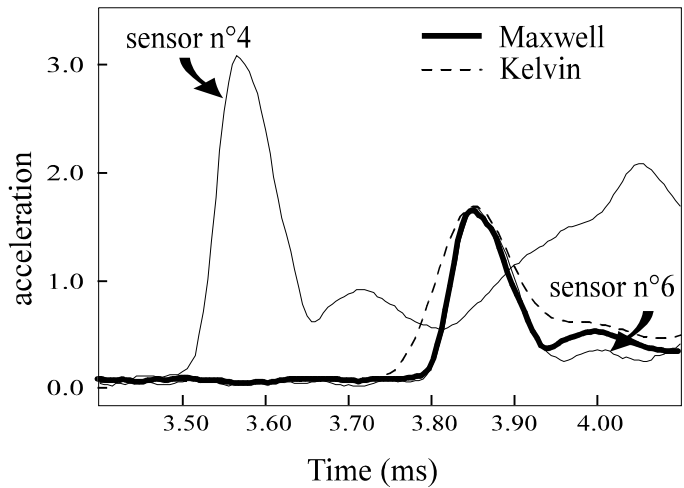

Figure 17 : *Calculated acceleration signals for different viscoelastic simulations : magnitudes in the XZ plane (drop-ball experiments).*

Acceleration signals obtained from these values of viscosity coefficient are presented on Figure 17. The acceleration given by sensor no.4 is the reference signal. Calculated accelerations at distance $r=r_6-r_4$ from reference sensor are compared with that given by sensor no.6. Judging by the curves of Figure 17, the Maxwell model seems to give simulated acceleration signals very close to the measured signals (for the chosen $\eta$ value). The results are very poor for the Kelvin model.

### 6.3.4 Optimisation procedure (Zener)

To determine the best behaviour parameters ($E_v$ and $\eta$) for the Zener model (see Figure 14), it is necessary to quantify the agreement between measured and simulated accelerations. It is, for instance, possible to estimate the squared error between measured signals and calculated signals. Considering the time interval $[t_1;t_2]$ around first acceleration peak, the functional $J(E_v,\eta)$, defining the squared error between measured acceleration $a_m(r_j)$ and calculated acceleration $a_m^{sim}(r_j)$ can be written :

$$J(E_v,\eta) = \frac{1}{J_0} \cdot \int_{t_1}^{t_2} \left( a_m(r_j,t) - a_m^{sim}(r_j,t) \right)^2 . dt$$

A minimization of $J(E_v,\eta)$ gives the optimum parameters $(E_{v,opt},\eta_{opt})$. The 3D curve of Figure 18 gives values of the functional $J(E_v,\eta)$ for $E_v \in [0; 50\ \text{MPa}]$ and $\eta \in [130; 160\ \text{kPa.s}]$ (for sensors 4 and 6).

The minimum error $J_{min}$ is obtained for $E_v$=25.5 MPa and $\eta$=146 kPa.s. The optimum "viscous" modulus $E_{v,opt}$ is very low, and the optimum viscosity coefficient $\eta_{opt}$ is very close to the viscosity value chosen for the Maxwell model. With these two optimum behaviour parameters, the simulations with the Zener and Maxwell models are very close. Both models have then to be compared from a quantitative point of view. Calculating the error $J(\eta)$ for the Maxwell model, the minimum value corresponds to an optimum viscosity $\eta_{opt}$=144 kPa.s. This value differs little from that determined empirically earlier ($\eta$=150 kPa.s). The minimum value of $J(\eta)$ for the Maxwell model is $J_{min}^{Max}$=7.02.10$^{-3}$. For the Zener model, it is a little smaller $J_{min}^{Zen}$=6.80.10$^{-3}$. The fit of the Zener model is slightly better than Maxwell's.

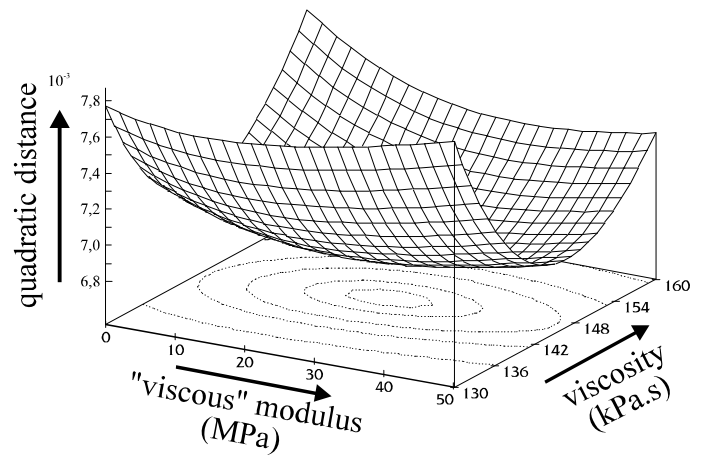

Figure 18 : *Determination of optimal behaviour parameters from error minimization (Zener model).*

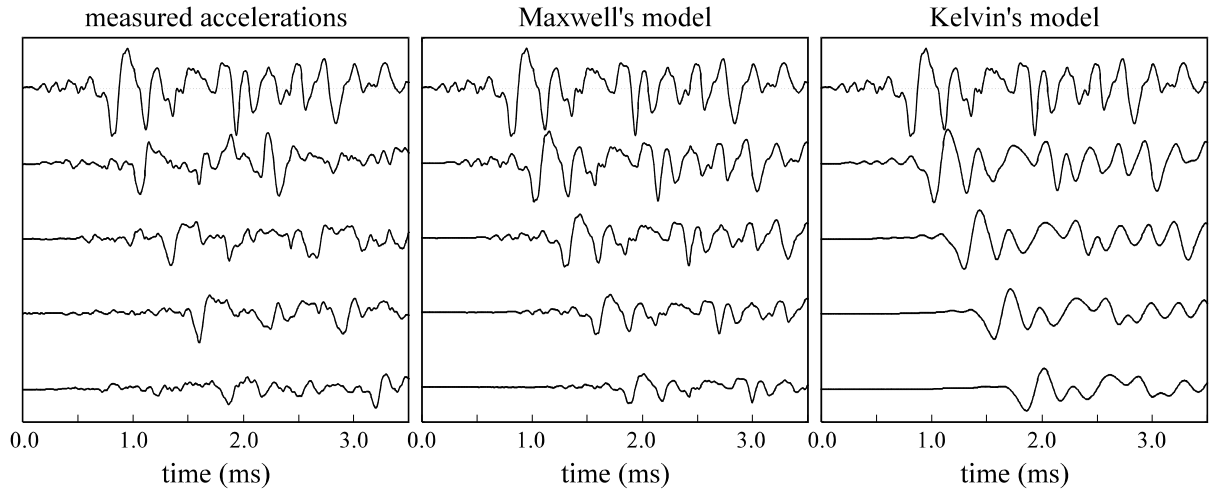

Figure 19 : *Calculated acceleration signals for different viscoelastic simulations (explosive earthquake simulations).*

### 6.4 Simulations for earthquake simulations

Using values of optimal viscosity estimated from drop-ball experiments ($\eta$=150 kPa.s for Maxwell's model, $\eta$=1000 Pa.s for Kelvin's model), we performed viscoelastic wave propagation simulations for explosive earthquake experiments. Figure 19 gives time acceleration curves for experimental measurements, Maxwell simulations and Kelvin simulations. For the left, center and right groups of curves, the first always corresponds to the same experimental acceleration measurement (sensor 2). From these curves, one can notice that damping phenomena are well estimated concerning acceleration amplitude. The form of the propagated waves is however more realistic for Maxwell's model, because there is a strong attenuation of high frequency components with Kelvin's model.

Since there is a strong soil-container interaction for explosive earthquake simulations, it is nevertheless difficult to compare directly experimental and numerical results. It is not really possible to compare the form and the amplitude of every acceleration peak because of the duration of the wave train (which implies wave reflections on medium boundaries) and the possible wave travel in the container itself.

### 6.5 Phase and group velocities

If $\xi(\omega)$ is the complex wave number, the classical expressions of phase and group velocities become complex. The real parts of these expressions correspond to phase shift terms and characterise the phase velocity $V_\Phi$ and group velocity $V_g$.
Their expressions then take the following form :

$$V_\Phi = Re\left[\frac{\omega}{\xi(\omega)}\right] \quad \text{and} \quad V_g = Re\left[\frac{d\omega}{d\xi(\omega)}\right]$$

where *Re* denotes the real part of a complex number.

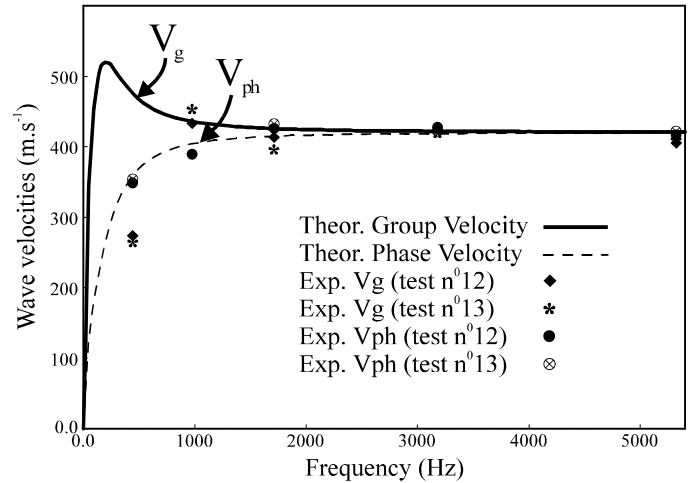

Figure 20 : *Comparison between theoretical phase and group velocities (from Maxwell viscoelastic simulations) and experimental ones (estimated from filtered signals).*

The determination of viscoelastic behaviour parameters allows the calculation of phase and group velocities for mid-depth sensors. The curves, given in Figure 20, indicate the values of $V_\Phi$ and $V_g$ for a Maxwell viscoelastic simulation. In Figure 20, these curves are compared with experimentally estimated values already given in Table IV. It allows a comparison in the frequency range [0 Hz; 5500 Hz]. Above this range, experimental and theoretical values coincide perfectly.

Between 0 and 5500 Hz, the comparison between experimental and theoretical values leads to the following conclusions :
- above *1500 Hz* : phase and group velocities are almost identical; there is a good agreement between experimental and theoretical values ;
- at *975 Hz* : group velocity is higher than phase velocity for theoretical values as well as for experimental values; the agreement between both results is not too bad ;
- at *440 Hz* : theoretical group velocity is higher than theoretical phase velocity; experimental phase velocities are not far from theoretical ones, but for group velocities there is a large discrepancy. Actually, wavelengths for low frequencies (*f<1000Hz*) are large and group velocity is greatly underestimated.

As indicated previously, there is a strong wave-guide effect at low frequencies. The underestimation of phase velocity is not significant, but for group velocity there is a strong influence of wave reflection (wave-guide effect). For low frequencies, geometrical dispersion has to be taken into account. This is the main drawback of dynamic experiments in a finite medium.

## 7. SIMULATIONS AT DIFFERENT SOURCE DISTANCES

Previous simulations were all made for the same pair of sensors (no.4 and no.6). It is necessary to make other simulations for different pairs of sensors to study the influence of sensor location and acceleration level. Determining the optimum parameters, as presented before, allows a comparison between all results.

For the Maxwell model, four pairs of "excitation-response" (in-out) signals are taken from the measured accelerations of sensors 2/4; 4/6; 6/8 and 8/10. The error term $J(\eta)$ is determined in each case to estimate the optimal viscosity value $\eta_{opt}$. As shown in Table VI, the results are very different from one case to another.

For the Maxwell model, the attenuation $Q^{-1}$ is related to behaviour parameters by the relation :

$$Q^{-1} = \frac{E}{\eta.\omega}$$

Using a linear viscoelastic model, it is possible to quantify damping phenomena. However, the proposed analytical description cannot describe the attenuation process in the whole medium. Some particular phenomena could have a strong influence on damping : amplitude dependence of attenuation, mean stress dependence. From the results given in Table VI, material attenuation seems to be very strong near the source of excitation (low viscosity, high acceleration magnitude) and much lower than geometrical attenuation for the last sensors (8 and 10), which reduces the acceleration amplitude by approximately 20 % (see Table V).

| sensor | | behaviour parameters | |
|---|---|---|---|
| in | out | viscosity | $J(\eta)$ |
| 2 | 4 | 39 kPa.s | $1,63.10^{-2}$ |
| 4 | 6 | 144 kPa.s | $7,02.10^{-3}$ |
| 6 | 8 | 154 kPa.s | $2,22.10^{-2}$ |
| 8 | 10 | +∞ no material attenuation | |

Table VI : *Optimal (Maxwell) viscosity values for simulations at different distances from the source.*

As attenuation could depend on acceleration amplitude [12] or mean stress, the need for a more complete approach is clear. However, the constitutive models presented in this paper give a complete analytical description of a spherical wave field propagation related to drop-ball experiments. The quantification of attenuation is therefore made very easy.

## 8. CONCLUSION

Wave propagation phenomena have been investigated for a medium of finite extent (centrifuge soil mass) by considering three-dimensional acceleration components, wave reflections at the boundaries (homomorphic filtering), time-frequency processing and dispersive phenomena, attenuation and energy dissipation (linear viscoelastic simulations).

An original experimental arrangement (drop-ball) is proposed. During these experiments, acceleration waves generated at mid-depth are shown to be pure pressure waves. Wave reflections at boundaries are detected and filtered using a special signal-processing method. Dispersive phenomena are investigated and are shown to take two different forms : strong geometrical dispersion at low frequencies and weaker material dispersion at higher frequencies. For mid-depth sensors, pressure wave propagation is completely described by spherical wave propagation in a viscoelastic medium. It allows the estimation of corresponding constitutive parameters and gives a quantification of dispersive and damping phenomena. We found a good numerical fitting when using rheological models but a realistic justification must be based on physical deformation mechanisms.